\documentstyle[aps]{revtex}

\begin{document}

\title{
  \begin{flushright} \begin{small}
     gr-qc/0009042
  \end{small} \end{flushright}
\vspace{1.cm}
Dyonic Wormholes in $5D$ Kaluza-Klein Theory
}
\author{Chiang-Mei Chen\footnote{E-mail: cmchen@phys.ntu.edu.tw}}
\address{Department of Physics,
         National Taiwan University, Taipei 106, Taiwan, R.O.C.}
\maketitle

\begin{abstract}
New spherically symmetric dyonic solutions, describing a
wormhole-like class of spacetime configurations in
five-dimensional Kaluza-Klein theory, are given in an explicit
form. For this type of solution the electric and magnetic fields
cause a significantly different global structure. For the electric
dominated case, the solution is everywhere regular but, when the
magnetic strength overcomes the electric contribution, the mouths
of the wormhole become singular points. When the electric and
magnetic charge parameters are identical, the throats
``degenerate'' and the solution reduces to the trivial embedding
of the four-dimensional massless Reissner-Nordstr{\"o}m black hole
solution. In addition, their counterparts in eleven-dimensional
supergravity are constructed by a non-trivial uplifting.
\end{abstract}

\bigskip
{PACS number(s): 04.20.Jb, 04.50.+h}
\bigskip


\section{Introduction}
After the brilliant insight of Kaluza and Klein \cite{KK}
realizing that the Einstein gravity and Maxwell electromagnetism
theories can be unified in a five-dimensional manifold, the
Kaluza-Klein theory, essentially the five-dimensional pure general
relativity, has developed explosively. (A comprehensive discussion
of Kaluza-Klein theory is given in a recent review article
\cite{OvWe97}.) This is the first theory advocating the idea that
the physical world may have more than four dimensions; it laid a
foundation for modern developments such as superstrings and
M-theory.

The exact solutions, especially spherically symmetric ones, of
Kaluza-Klein theory have been studied extensively. A subset of
such solutions corresponds to black holes. Well-known examples
include the electric pp-wave obtained by Dobiasch and Maison
\cite{DoMa82} and the magnetic GPS monopole of Gross, Perry
\cite{GrPe83} and Sorkin \cite{So83}. Later, Gibbons and Wiltshire
\cite{GiWi86} successfully unified the pp-wave and GPS monopole
into a dyonic solution which recently has been generalized, by
Rasheed \cite{Ra95}, to include a rotation. Furthermore, the
hidden symmetry, $SL(3,R)$, of the vacuum $5D$ metric with two
commuting isometries and its solution generating application were
analysed in \cite{Ma79,Cl86a,Cl86b} and were extended in
\cite{CvYo95}.

In addition to the black holes, there is another subset of
solutions, the so-called {\em wormholes}, which were introduced
originally by Wheeler \cite{Wh55}. The most physical interesting
property of wormholes is that this type of solution provides a
possible, theoretically at the moment, way for time traveling,
which, if it can be realized, may lead to a break down of some
traditional concepts of nature, especially causality
\cite{MoThYu88} (see \cite{Vi95} and references therein).

The discovery of wormholes for Kaluza-Klein theory dated back to
the work done by Chodos and Detweiler \cite{ChDe82}. In that
paper, a class of regular, spherically symmetric and
asymptotically flat solutions characterized by three parameters
(mass, electric and scalar charges) was constructed and certain
cases were interpreted as wormholes. Afterwards, this class of
solutions was generalized to axisymmetric multi-wormholes
\cite{Cl84a} and to higher dimensions \cite{Cl84b} by Cl\'ement.
More recently, a class of wormholes with diagonal $5D$ metric was
also discussed in \cite{AgBiLiWe99}.

The purpose of present paper is giving a dyonic extension of the
massless electric wormhole solutions given in
\cite{ChDe82,Cl84a,Cl84b} and rediscovered recently by
Dzhunushaliev \cite{Dz98}. Surprisingly, the exact form of these
dyonic wormholes is elegant and succinct. The primary properties
of these dyonic wormholes are discussed. In addition, their
eleven-dimensional supergravity counterparts are presented.

\section{Kaluza-Klein Dyonic Wormholes}
For a general spherically symmetric dyon characterized by two
parameters $q$ (electric) and $p$ (magnetic), we take the following
general form for the metric
\begin{equation}
ds_5^2 = \frac{B}{A} \left( d\chi + \omega dt + 2p\cos\theta
d\varphi \right)^2 - \frac{\Delta}{B} dt^2 + A \left(
\frac{dr^2}{\Delta} + d\Omega^2 \right), \label{DA}
\end{equation}
where $\chi$ is the extra fifth coordinate, $d\Omega^2 :=
d\theta^2 + \sin^2\theta d\varphi^2$ and $A,B,\omega,\Delta$ are
functions which depend only on the variable $r$. The value of
coordinate $r$ is extended to $\{-\infty, \infty\}$.

Regarding the wormhole solutions, the pure electric case found in
\cite{ChDe82,Cl84a,Cl84b,Dz98} can be rewritten, by choosing
appropriate coordinates and discarding a dummy parameter, in a
more succinct form which satisfies the Eq. (\ref{DA}) pattern
together with the condition $p=0$ and the following specific
functions
\begin{equation}
\hbox{\bf electric:} \qquad \omega_e = \frac{2qr}{r^2-q^2}, \qquad
A_e = \Delta_e = r^2+q^2, \qquad B_e = r^2-q^2. \label{EWH}
\end{equation}
Hereafter, we use the subscripts $e,m$ and $d$ to denote the
electric, magnetic and dyonic solutions respectively.

Obviously, the likely ``singularities'' of above spacetime
configuration (\ref{EWH}) can locate at $r^2=q^2$ which divide the
entire spacetime into three different regions: two {\em outer
regions} $r^2>q^2$ (two slices, $r<-|q|$ and $r>|q|$) and one {\em
core region} $r^2<q^2$, ($-|q|<r<|q|$). Furthermore, it is easy to
recognize that in the two outer regions are asymptotically flat,
but in the core region, the $t$ coordinate changes its sign and
becomes space-like \cite{DzSc99} while the elsewhere space-like
fifth coordinate $\chi$ changes to time-like. It was shown in
\cite{ChDe82} that the solution (\ref{EWH}) is regular everywhere
for $r\in\{ -\infty,\infty\}$, so that the five-dimensional
geometry is of the Lorentzian wormhole type. However, as pointed
out in \cite{AzCl90} this wormhole, with only a pure electric
charge, is non-traversable in the sense that a physical
(non-tachyonic) test particle cannot go from one asymptotic flat
region ($r\to\infty$) to the other ($r\to-\infty$). The ``fake''
singularities located at $r^2=q^2$ are just two mouths of the
throat of the wormhole and the area of each mouth is finite, $4
\pi q^2$. The detailed analysis, including geometrical properties
and stability of these solutions, has been given in \cite{AzCl00}.

Unfortunately, the exact magnetic solution is ``unknown'' in the
literature, instead of which Dzhunushaliev and Singleton gave a
numerical analysis \cite{DzSi99b} to discuss its expected
behavior. The authors claimed that this magnetic solution is a
finite {\em flux tube}, which may provide a reason why free
monpoles do not appear to exist in nature: they are confined into
monopole-anti monopole pairs in a finite, flux tube-like spacetime
that is similar to the flux tube confinement picture of quarks in
QCD.

However, every Kaluza-Klein magnetic solution, including the GPS
monopole and, of course, this desired new type, can be derived by
$S$-duality from its electric dual partner. (There is a lot of
literature discussing this duality, see e.g. \cite{ChGaMaSh99}).
Applying $S$-duality, the ``dual'' solution of (\ref{EWH}) can be
found easily without solving the field equations. Its exact form
is of the form (\ref{DA}) along with $q=0$ and
\begin{equation}
\hbox{\bf magnetic:} \qquad \omega_m = 0, \qquad A_m = r^2-p^2,
\qquad B_m = \Delta_m = r^2+p^2. \label{MWH}
\end{equation}

According to (\ref{MWH}), the possible singularities could occur
at $r^2=p^2$ and the two outer regions $r^2>p^2$ are also flat
asymptotically. However, in the core region $r^2<p^2$ all
space-like coordinates change their sign and the signature of this
region is ``all minus'' which indicates that the spacetime is not
Lorentzian anymore but pseudo-Euclidean. Nevertheless, an
apparently wormhole-like (pseudo-Euclidean) configuration, but not
a flux tube as claimed in \cite{DzSi99b}, may again be formed in
the magnetic solution. However, the geometrical structure of these
magnetic solutions is significantly different from the electric
ones. The two-surfaces of each ``end'', located at $r=\pm p$, of
the core region for these magnetic solutions, covered by the
coordinates $\theta$ and $\varphi$, shrink to a point for all
values of the magnetic charge. We have checked that the magnetic
solutions are indeed singular at $r=\pm p$ (the $5D$ Kretschmann
invariant, $K_5=R_{\alpha\beta\mu\nu} R^{\alpha\beta\mu\nu}$, is
divergent at these points), so that these solution actually do not
form wormholes since their spacetime can not be extended
analytically from $r \to -\infty$ to $r \to \infty$. In general,
with more than one time-like dimension, one opens up the
possibility of closed time-like curves and causality violations.
However, since the region of the ``throat'' seems to be ``pinched
off'' from the asymptotic regions it probably does not cause any
mischief.

It is worth noting that our purely magnetic solution is just the
Euclideanized massless Taub-NUT solution \cite{HaEl73} with a
trivial time direction\footnote{I am grateful to a referee for
pointing out this analogue.}.

The next task, naturally, is trying to construct a two-parameter
dyon which can combine the above two single-parameter electric
wormhole and magnetic singular solutions. With the help of the
symbolic calculation package {\sl GRG} \cite{Ze97}, we have
obtained this kind of dyonic solution. The result, written in the
form of Eq. (\ref{DA}), is
\begin{equation}
\hbox{\bf dyonic:} \qquad \omega_d = \frac{2qr}{r^2+p^2-q^2},
\quad A_d = r^2-p^2+q^2, \quad B_d = r^2+p^2-q^2, \quad \Delta_d =
r^2+p^2+q^2. \label{DWH}
\end{equation}

We now summarize the essential properties, from the
five-dimensional point of view, of these new dyons:

\begin{itemize}

\item The solutions (\ref{DWH}) have a remarkably elegant and symmetric
expression, even simpler than the well-known dyonic black holes in
\cite{GiWi86}.

\item For the {\em electric dominated} case, $q^2>p^2$, this
solution is regular everywhere, which can primarily be verified by
the associated Kretschmann invariant.\footnote{A more rigorous way
to observe the regularity is by the fact that the determinant of
the five-dimensional metric, $|g_5|=A_d^2\sin^2\theta$, is
non-vanishing for all values of $r$ when $q^2>p^2$. Therefore, one
can, for instance, everywhere locally transform to a frame where
the metric tensor is diagonal with finite and non-vanishing
elements.} With extended value of $r \in \{-\infty,\infty\}$, it
describes the configuration of a Lorentzian wormhole connecting
two asymptotically flat spacetimes. The area of the mouths is
finite: $4\pi(q^2-p^2)$.

\item For the {\em magnetic dominated} case, $p^2>q^2$, the core region
is pseudo-Euclidean and the ``mouths'' at $r=\pm\sqrt{p^2-q^2}$
compress to singular points. Actually, these two points are
singular thus the geometry can not be extended analytically from
one Lorentzian asymptotical flat spacetime to the other through
the core.

\item For the case $q=\pm p$, the throat degenerates to a point at $r=0$
and the Kaluza-Klein dyonic solutions are just the trivial
embedding of the four-dimensional massless Reissner-Nordstr{\"o}m
black hole solutions of the Einstein-Maxwell theory.

\end{itemize}

\section{Four-Dimensional Observer's Examination}
There is a physically important question to ask: What will be
experienced for a four-dimensional observer inhabiting the
spacetime described by (\ref{DWH})? In order to be consistent with
Einstein's theory of gravity, the four-dimensional effective
action reduced from the five-dimensional general relativity should
be presented in the Einstein frame forming the so-called
Einstein-Maxwell-dilaton theory with a certain dilaton coupling
constant (minimal coupling of the scalar field). The $4D$ metric
of (\ref{DA}) in the Einstein frame is given by
\begin{equation}
d s_4^2 = - \frac{\Delta}{\sqrt{AB}} dt^2 + \sqrt{AB} \left(
  \frac{dr^2}{\Delta} + d\Omega^2 \right). \label{EM4D}
\end{equation}

Therefore, in the Einstein frame, the reduced metric (\ref{EM4D})
of dyonic solutions (\ref{DWH}) shows:

\begin{itemize}

\item The distinctness between electric, $q$, and magnetic, $p$,
parameters disappears. (Since $\Delta_d=r^2+p^2+q^2$ and
$\sqrt{A_d B_d}=\sqrt{r^4-(p^2-q^2)^2}$, thus the metric
(\ref{EM4D}) is invariant under exchange parameters $p$ and $q$
for dyonic solutions (\ref{DWH}).)

\item The asymptotical limit, $r^2\to\infty$, is just the dyonic
massless Reissner-Nordstr{\"o}m solution,
\begin{equation}
d s_4^2 \sim - \left( 1+\frac{p^2+q^2}{r^2} \right) dt^2 + \left(
1 + \frac{p^2+q^2}{r^2} \right)^{-1} dr^2 + r^2 d\Omega.
\end{equation}

\item The two-surface area of ``origin'' for the electromagnetic charges at
$r^2=|p^2-q^2|$ is always zero.

\item The core region $r^2<|p^2-q^2|$ physically does not exist, or
it is ``invisible'', due to the fact that $\sqrt{A_d B_d}$ is
ill-defined in this region. From the dimensional reduction point
of view, in this region, the KK reduction along the direction
$\partial_\chi$ is not valid since the Killing vector becomes
time-like and presumably does not have closed orbits.

\end{itemize}

Synthesizing the above arguments, one can claim that within the
dyonic solutions (\ref{DWH}) a four-dimensional observer will
``see'' only one of the two regions of spacetime generated by
electromagnetic ``point-like'' sources. Nevertheless, for the
four-dimensional observer these charges are entirely
``disconnected''. From the five-dimensional point of view,
however, the explanation may be complete different --- the charges
could be connected by a wormhole. Thus, our solutions provide a
possible realization of the model of ``charge without charge''
idea for electromagnetic sources proposed by Wheeler.

\section{Counterparts in Eleven-Dimensional Supergravity}
One can show \cite{ChGaSh00,ChGaSh99} that there exists a
``duality'' between eight-dimensional vacuum configurations
possessing two commuting space-like Killing vectors and
eleven-dimensional supergravity solutions satisfying a certain
ansatz. Applying this correspondence to our general ansatz of
metric (\ref{DA}) smeared to eight dimensions, one can obtain the
related counterparts within the framework of M-theory. The
solutions are
\begin{eqnarray}
ds_{11}^2 &=& \left(\frac{B}{A}\right)^{-2/3} \left(dx_1^2 +
dx_2^2\right) + \left(\frac{A}{B}\right)^{-1/3} \left(
\sum_{i=3}^7 dx_i^2 \right) + \left(\frac{B}{A}\right)^{1/3}
\left[ -\frac{\Delta}{B}dt^2 + A \left( \frac{dr^2}{\Delta} +
d\Omega^2 \right) \right], \\
\hat A_{tx_1x_2} &=& \omega, \qquad \hat A_{\varphi x_1x_2} = 2 p
\cos\theta.
\end{eqnarray}
Considering only the pure electric solutions, the
eleven-dimensional metric and form field reduce to
\begin{eqnarray}
ds_{11}^2 &=& \left(\frac{r^2-q^2}{r^2+q^2}\right)^{-2/3} \left(
-dt^2 + dx_1^2 + dx_2^2 \right) + \left( \frac{r^2-q^2}{r^2+q^2}
\right)^{1/3} \left[ dr^2 + (r^2+q^2) d\Omega^2 + \sum_{i=3}^7
dx_i^2 \right], \\
\hat A_{tx_1x_2} &=& \frac{2qr}{r^2-q^2}.
\end{eqnarray}
Thus, this is another type of $2$-brane which is different from
the already known $M2$-brane and $M2$-fluxbrane to the
eleven-dimensional supergravity. The proper terminology for the
above solution intuitively should be ``{\em $M2$-wormbrane}''.
Similarly, the counterparts of the magnetic and dyonic solutions
are a new type of $5$-brane and $2\cup 5$-brane
\cite{ChGaSh00,ChGaSh99}. What physical role may be played by
these solutions is still unclear and needs further investigation.

\section{Conclusion}
In this paper, we obtain new spherically symmetric dyonic
solutions including wormholes and configurations with naked
singularities in the five-dimensional Kaluza-Klein theory. The
geometrical structure is determined by the relative strength of
the electric and magnetic charges. When the electric charge
dominates, i.e. $q^2>p^2$, the solution is a Lorentzian wormhole
and the mouths of the wormhole have finite area. Where, when the
magnetic charge dominates, $p^2>q^2$, the core region is
pseudo-Euclidean and its ends shrink to singular points. For this
case, the spacetime can not be extended analytically from $r \to
-\infty$ to $r \to \infty$. Moreover, if the electric and magnetic
fields are in ``balance'', $p^2=q^2$, the core degenerates to a
point and the solution is just a trivial embedding of $4D$
massless Reissner-Nordstr{\"o}m black holes.

However, a four-dimensional observer cannot detect the existence
of the core region but rather sees a spacetime generated by a
point-like electromagnetic source. Therefore, these wormhole-like
solutions show that the Wheeler's ``charge without charge'' for
the origin of electric and magnetic charges may can be realized in
higher dimensions.

It is worth noting that there is an one-parameter flux tube
solution which was given in \cite{DzSi99a,DzSi99b}, explicitly
\begin{equation}
\hbox{\bf Flux tube:} \qquad \omega_F = \frac{r}{p}, \qquad
  A_F = B_F = 2p^2, \qquad \Delta_F = r^2 + 2p^2. \label{FT}
\end{equation}
This solution was expected to be an extreme limit of a {\it dyonic
solution} which ``can'' combine an electric wormhole and a
magnetic flux tube. However, in this paper we have shown that the
dyonic extension from the electric wormhole couples to the
magnetic singular solutions but there is no flux tube. Moreover,
the solution (\ref{FT}) can easily be understood as belonging to
another category of solutions since it is not asymptotic flat.
Therefore, there should exist a generalized {\it dyonic flux tube}
the extreme case of which is just the solution (\ref{FT}).

\section*{Acknowledgments}
The author would like to thanks G. Cl\'ement, V. Dzhunushaliev,
D.V. Gal'tsov, J.M. Nester and D. Singleton for many stimulating
discussions and helpful comments in preparing this article. This
work is supported by the Taiwan CosPA project.


\end{document}